\documentclass[onecolumn]{revtex4}

\usepackage{graphicx}% Include figure files
\usepackage{dcolumn}% Align table columns on decimal point
\usepackage{bm}% bold math

\input epsf
\begin{document}

\title{\Large{Correspondence between Electro-Magnetic Field and other
Dark Energies in Non-linear Electrodynamics}}

\author{\bf Sayani Maity$^1$\footnote{sayani.maity88@gmail.com},
Shuvendu Chakraborty$^2$\footnote{shuvendu.chakraborty@gmail.com}
and Ujjal Debnath$^1$\footnote{ujjaldebnath@yahoo.com,
ujjal@iucaa.ernet.in}}

\affiliation{$^1${Department of Mathematics, Bengal Engineering
and Science University, Shibpur, Howrah-711 103, India.\\
$^2$Department of Mathematics, Seacom Engineering College, Howrah,
711 302, India.}}

\date{\today}

\begin{abstract}
In this work, we have considered the flat FRW model of the
universe filled with electro-magnetic field. First, the Maxwell's
electro-magnetic field in linear form has been discussed and after
that the modified Lagrangian in non-linear form for accelerated
universe has been considered. The corresponding energy density and
pressure for non-linear electro-magnetic field have been
calculated. We have found the condition such that the
electro-magnetic field generates dark energy. The correspondence
between the electro-magnetic field and the other dark energy
candidates namely tachyonic field, DBI-essence, Chaplygin gas,
hessence dark energy, k-essenece and dilaton dark energy have been
investigated. We have also reconstructed the potential functions
and the scalar fields in this scenario.
\end{abstract}

\maketitle

\section{\normalsize\bf{Introduction}}

Recent observations of type Ia supernovae (SNIa) indicate that our
universe is now undergoing an accelerating expansion [1]. The main
responsible candidate for the cosmic acceleration is generally
dubbed as `dark energy', a mysterious exotic energy with negative
pressure. Present cosmological observational data suggest that
universe is dominated by this dark energy with 70 percent of the
total. So the feature of the universe naturally depends on the
nature of the dark energy. the simplest candidate of dark energy
is Cosmological Constant with fixed equation of state (EOS)
$w=-1$. If it is quintessence then $-1<w<-1/3$ and if it is
phantom then $w<-1$. The constant EOS $w=-1$ is called phantom
divide. There are some dark energies which can cross the
phantom divide from both sides.\\

In present years the standard cosmological model based on
Friedmann-Robertson-Walker (FRW) with Maxwell electrodynamics has
got much attention and many interesting results are obtained. This
leads to a cosmological singularity at a finite time in the past
and result the energy density and curvature arbitrary large in the
very early epoch [2]. This singularity breaks the laws of physics
with mathematical inconsistency and physical incompleteness of any
cosmological model. There are some proposals to handle this
primordial singularity such as cosmological constant [3], non
minimal couplings [4], modifications of geometric structure of
space-time [5], non-equilibrium thermodynamics [6], Born-Infeld
type nonlinear electromagnetic field [7] and so on. Recently a new
approach has been taken to avoid the cosmic singularity through a
nonlinear extension of the Maxwell electromagnetic theory. Another
interesting feature can be viewed that an exact regular black hole
solution has been recently obtained proposing Einstein-dual
nonlinear electrodynamics [8, 9]. Exact solutions of the
Einstein's field equations coupled with nonlinear electrodynamics
(NLED) reveal an acceptable nonlinear effect in strong
gravitational and magnetic fields. Also the General Relativity
(GR) coupled with NLED effects can explain the primordial
inflation.\\

In this work, we have briefly discussed the Maxwell's
electrodynamics in linear and non-linear forms and then the
modified Lagrangian in accelerated universe has been considered.
The energy density and pressure for non-linear electro-magnetic
thery have been calculated. The purpose of the present work is to
investigate the correspondence between electromagnetic field and
other dark energy candidates namely tachyonic field [10],
DBI-essence [11], Chaplygin gas [12], hessence dark energy [13],
k-essenece [14] and dilaton dark energy [15]. The potentials have been
reconstructed in terms of electric field and magnetic field in above types of
dark energy models.\\

\section{\normalsize\bf{Electro-magnetic Theory in Non-linear Electrodynamics}}

Lagrangian density in Maxwell's electrodynamics can be written as
[16]

\begin{equation}
{\cal
L}=-\frac{1}{4\mu_{0}}~F^{\mu\nu}F_{\mu\nu}=-\frac{1}{4\mu_{0}}~F
\end{equation}

where $F^{\mu\nu}$ is the electromagnetic field strength tensor
and $\mu_{0}$ is the magnetic permeability. The canonical
energy-momentum tensor is then given by

\begin{equation}
T_{\mu\nu}=\frac{1}{\mu_{0}}\left(F_{\mu\alpha}F^{\alpha}_{\nu}+\frac{1}{4}~Fg_{\mu\nu}
\right)
\end{equation}

Since the spatial section of FRW geometry are isotropic,
electromagnetic fields can generate such a universe only if an
averaging procedure is performed [17]. Applying standard spatial
averaging process for electric field $E_{i}$ and magnetic field
$B_{i}$, set

\begin{equation}
<E_{i}>=0,~~<B_{i}>=0,~~<E_{i}E_{j}>=-\frac{1}{3}~E^{2}g_{ij},
~~<B_{i}B_{j}>=-\frac{1}{3}~B^{2}g_{ij},~~<E_{i}B_{j}>=0.
\end{equation}

So one gets,

\begin{equation}
<F_{\mu\alpha}F^{\alpha}_{\nu}>=\frac{2}{3}\left(\epsilon_{0}E^{2}+
\frac{B^{2}}{\mu_{0}}
\right)u_{\mu}u_{\nu}+\frac{1}{3}\left(\epsilon_{0}E^{2}-
\frac{B^{2}}{\mu_{0}} \right)g_{\mu\nu}
\end{equation}

where $u_{\mu}$ is the fluid 4-velocity. Now comparing with the
average value of energy momentum tensor

\begin{equation}
 <T_{\mu\nu}>=(\rho+p)u_{\mu}u_{\nu}-pg_{\mu\nu}~,
\end{equation}

the energy density and pressure have the forms

\begin{equation}
\rho=\frac{1}{2}\left(\epsilon_{0}E^{2}+ \frac{B^{2}}{\mu_{0}}
\right)~,~~~~p=\frac{1}{3}~\rho
\end{equation}

This implies the Maxwell's electro-magnetic fields generate
effectively the radiation fluid.\\

The modified Lagrangian in non-linear electrodynamics for
accelerated universe is considered as [18]

\begin{equation}
{\cal L}=-\frac{1}{4}~F+\alpha F^{2}+\beta F^{-1}
\end{equation}

where $\alpha$ and $\beta$ are arbitrary (constant) parameters. As
seen this Lagrangian contains both positive and negative powers of
$F$. The second (quadratic) term dominates during very early
epochs of the cosmic dynamics, while the Maxwell term (first term
above) dominates in the radiation era. The last term is
responsible for the accelerated phase of the cosmic evolution
[19]. The above Lagrangian density yields a unified scenario to
describe both the acceleration of the universe (for weak fields)
and the avoidance of the initial singularity, as a consequence of
its properties in the strong-field regime.\\

The energy density and pressure for electro-magnetic (EM) field
are given by

\begin{equation}
\rho_{B}=-{\cal L}-4E^{2}{\cal L}_{F}
\end{equation}
and
\begin{equation}
p_{B}={\cal L}-\frac{4}{3}(2B^{2}-E^{2}){\cal L}_{F}
\end{equation}

Now, the electro-magnetic field has the expression
$F=2(B^{2}-E^{2})$, so using (7) - (9) we get

\begin{equation}
\rho_{B}=\frac{1}{2}(B^{2}+E^{2})-4\alpha(B^{2}-E^{2})(B^{2}+3E^{2})-\frac{\beta}{2(B^{2}-E^{2})}
\end{equation}
and
\begin{equation}
p_{B}=\frac{1}{6}(B^{2}+E^{2})-\frac{4\alpha}{3}(B^{2}-E^{2})(5B^{2}-E^{2})+\frac{\beta(7B^{2}-5E^{2})}{6(B^{2}-E^{2})^{2}}
\end{equation}

The electro-magnetic field generates dark energy if the strong
energy condition is violated i.e., $\rho_{B}+p_{B}<0$. So from
(10) and (11), we get

\begin{equation}
(B^{2}+E^{2})-8\alpha(B^{2}-E^{2})(3B^{2}+E^{2})+\frac{\beta(3B^{2}-2E^{2})}{(B^{2}-E^{2})^{2}}<0
\end{equation}

In particular, if $\alpha=0$ and $\beta=0$, then the expression in
the l.h.s of (12) is $(B^{2}+E^{2})$ which cannot be negative. So
only Maxwell's electro-magnetic field (linear) cannot generate the
dark energy. So for getting dark energy, we require non-linear
electro-magnetic field.\\

The metric of a homogeneous and isotropic flat universe in the FRW
model is

\begin{equation} ds^{2}=-dt^{2}+a^{2}(t)\left[dr^{2}+
r^{2}(d\theta^{2}+\sin^{2}\theta d\phi^{2})\right]
\end{equation}

where $a(t)$ is the scale factor and $k ~(= ~0,~ +1,~ -1)$ is the
curvature scalar. The Einstein field equations for
electro-magnetic Universe are

\begin{equation}
3\frac{\dot{a}^{2}}{a^{2}}= \rho_{B}
\end{equation}
and
\begin{equation}
\frac{\ddot{a}}{a}= -\frac{1}{6}(\rho_{B}+p_{B})
\end{equation}
where $\rho_{B}$ and $p_{B}$ are energy density and pressure
corresponding to electromagnetic field given by the equations (10)
and (11) (choosing $8\pi G=c=1$). Now the energy-conservation
equation for electro-magnetic field is given by

\begin{equation}
\dot{\rho}_{B}+3\frac{\dot{a}}{a}(\rho_{B}+p_{B})=0
\end{equation}

Since non-linear electro-magnetic field generates dark energy, so
in the following sections, we shall discuss the correspondence
between the electro-magnetic (EM) field and the other types of
dark energies like tachyonic field, DBI-essence, Chaplygin gas,
hessence, k-essence and dilaton dark energies and hence find the
nature of the potentials and scalar fields.\\

\section{\bf{Correspondence between EM field and Tachyonic field}}

The present section aims to investigate the conditions under which
there is a correspondence between EM field and the tachyonic
field, in the flat FRW Universe. That is, to determine an
appropriate potential for tachyonic field which makes the two dark
energies to coincide with each other. Let us first consider the
energy density $\rho_{T}$ and the pressure $p_{T}$ for the
tachyonic field as [10]

\begin{equation}
\rho_{T}=\frac{V(\phi)}{\sqrt{1- \dot{\phi}^{2}}}
\end{equation}
and
\begin{equation}
p_{T}=-V(\phi)\sqrt{1- \dot{\phi}^{2}}
\end{equation}
where $\phi$ is the tachyonic field and $V(\phi)$ is the
corresponding potential. Comparing the energy density and pressure
corresponding to EM field and tachyonic field we have the
expressions for the tachyonic field and the potential as

\begin{equation}
\phi=\int
2\sqrt{\frac{(B^{2}-E^{2})^{2}(B^{2}+E^{2}+8(B^{4}-6B^{2}E^{2}+5E^{4})\alpha)-5B^{2}\beta+4E^{2}\beta}{
a^{2}(-B^{4}+E^{4}+8(B^{2}-E^{2})^{2}(B^{2}+3E^{2})\alpha+\beta)^{2}}}
da
\end{equation}

and

\begin{eqnarray*}
V=\frac{1}{2\sqrt{3}}\left[\left(B^{2}+E^{2}-40B^{4}\alpha +48
B^{2}E^{2}\alpha-8E^{4}\alpha+\frac{(7B^{2}-5E^{2})\beta}{(B^{2}-E^{2})^{2}}\right)\times
\right.
\end{eqnarray*}
\begin{equation}
\left.\left(-B^{2}-E^{2}+8(B^{4}+2B^{2}E^{2}-3E^{4})\alpha+\frac{\beta}{B^{2}-E^{2}}\right)\right]^{\frac{1}{2}}
\end{equation}

In figure 1, we have drawn the potential function $V$ against
magnetic field $B$ and electric field $E$. From 3 dimensional
figure, we have seen that $V$ always increases with $E$ increases
but slightly increases with increase in $B$.\\

\begin{figure}
\includegraphics[scale=0.6]{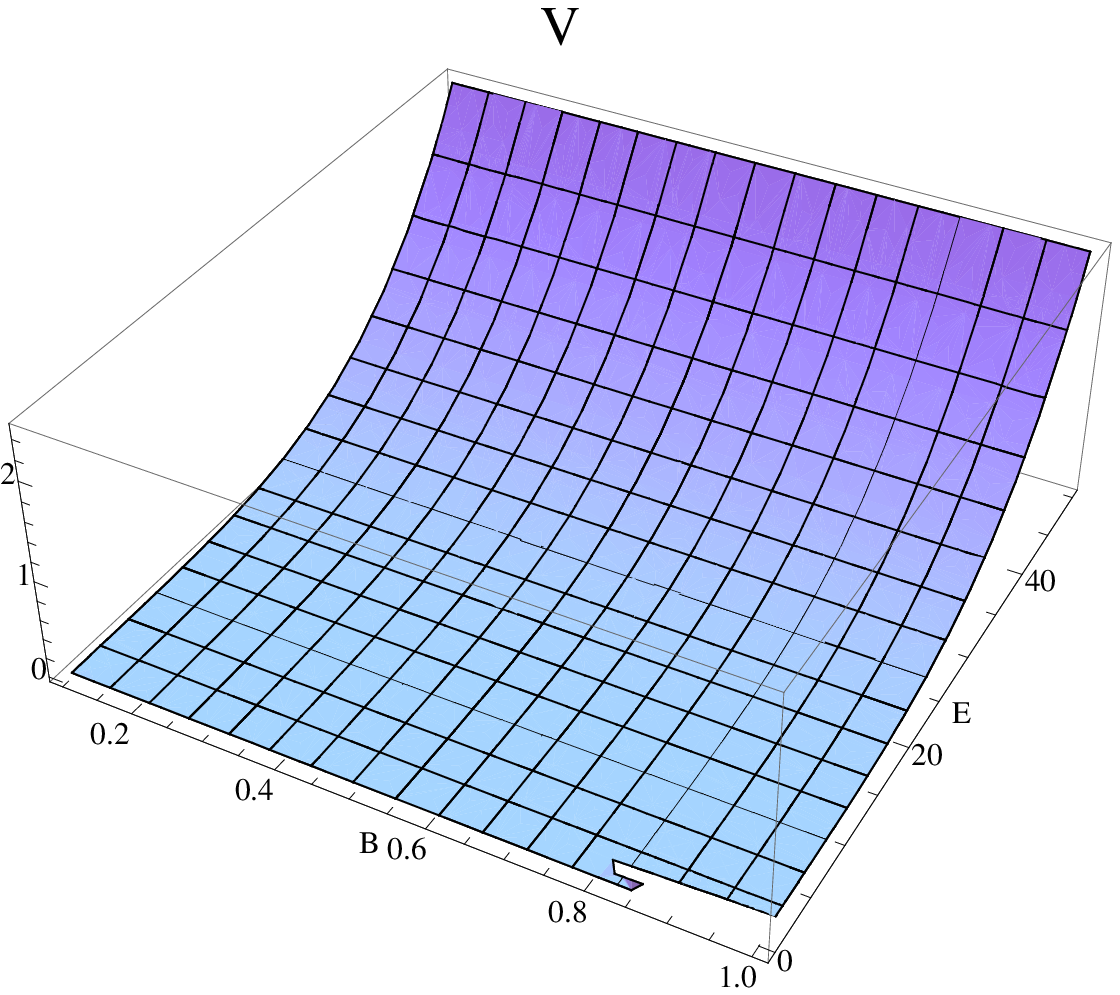}\\
\vspace{1mm} ~~~~~~~~~~~~Fig.1~~~~~~~~\\

\vspace{6mm} Fig. 1 shows the variation of $V$ against $B$ and $E$
for different values of $~\alpha=1,\,\beta=2$ when a
correspondence between EM field and tachyonic field is considered.

 \vspace{6mm}

\end{figure}

\section{\bf{Correspondence between EM field and DBI-essence}}
There have been many works aimed at connecting the string theory
with inflation. While doing so, various ideas in string theory
based on the concept of branes have proved themselves fruitful.
One area which has been well explored in recent years, is
inflation driven by the open string sector through dynamical
Dp-branes. This is the so-called DBI (Dirac-Born-Infield)
inflation, which lies in a special class of K-inflation models.
The energy density and pressure of the DBI-essence scalar field
are respectively given by [11]
\begin{equation}
\rho_{D}=(\gamma-1)T(\phi)+V(\phi)
\end{equation}
and
\begin{equation}
p_{D}=\frac{\gamma-1}{\gamma}T(\phi)-V(\phi)
\end{equation}

where $\gamma$ is given by
\begin{equation}
\gamma=\frac{1}{\sqrt{1-\frac{\dot{\phi}^{2}}{T(\phi)}}}
\end{equation}
where $V(\phi)$ is the self interacting potential and $T(\phi)$ is
the warped brane tension.\\

Comparing the energy density and pressure corresponding to EM
 field and DBI-essence scalar field we have the expressions for potential,
 wrapped brane tension and $\gamma$ for
 DBI-essence as
\begin{equation}
V(\phi)=\frac{1}{\gamma+1}(\rho_{D}-\gamma p_{D})
\end{equation}
and
\begin{equation}
T(\phi)=\frac{\rho_{D}+p_{D}}{\frac{\gamma^{2}-1}{\gamma}}
\end{equation}

where

\begin{equation}
\gamma=\frac{2B^{6}-2B^{4}E^{2}-2B^{2}E^{4}+2E^{6}-32B^{8}\alpha+64B^{6}E^{2}\alpha-64B^{2}E^{6}\alpha+
32E^{8}\alpha+2B^{2}\beta-E^{2}\beta}{3(B^{2}-E^{2})^{2}\dot\phi^{2}}
\end{equation}

Here we see that $V(\phi)$, $T(\phi)$ and $\gamma$ are functions
of $B^{2}, E^{2}$ and $\dot\phi^{2}$. To overcome the complexity
we consider here two cases $\gamma =$ constant and $\gamma \neq$
constant [20].\\

{\bf Case I:}  $\gamma$ = constant.\\

In this case, for simplicity, we assume $T(\phi)= n
\dot{\phi}^{2}$ $(n
> 1)$ and $V(\phi)= m \dot{\phi}^{2}$ ($m>0$). So we have $\gamma = \sqrt{\frac{n}{n-1}}$.

In this case the expression for $V(\phi)$ and $T(\phi)$ are given
by

\begin{equation}
V=m
\frac{2B^{6}-2B^{4}E^{2}-2B^{2}E^{4}+2E^{6}-32B^{8}\alpha+64B^{6}E^{2}\alpha-64B^{2}E^{6}\alpha+
32E^{8}\alpha+2B^{2}\beta-E^{2}\beta}{3(B^{2}-E^{2})^{2}\sqrt{\frac{n}{n-1}}}
\end{equation}

\begin{equation}
T=n
\frac{2B^{6}-2B^{4}E^{2}-2B^{2}E^{4}+2E^{6}-32B^{8}\alpha+64B^{6}E^{2}\alpha-64B^{2}E^{6}\alpha+
32E^{8}\alpha+2B^{2}\beta-E^{2}\beta}{3(B^{2}-E^{2})^{2}\sqrt{\frac{n}{n-1}}}
\end{equation}

\begin{figure}
\includegraphics[scale=0.6]{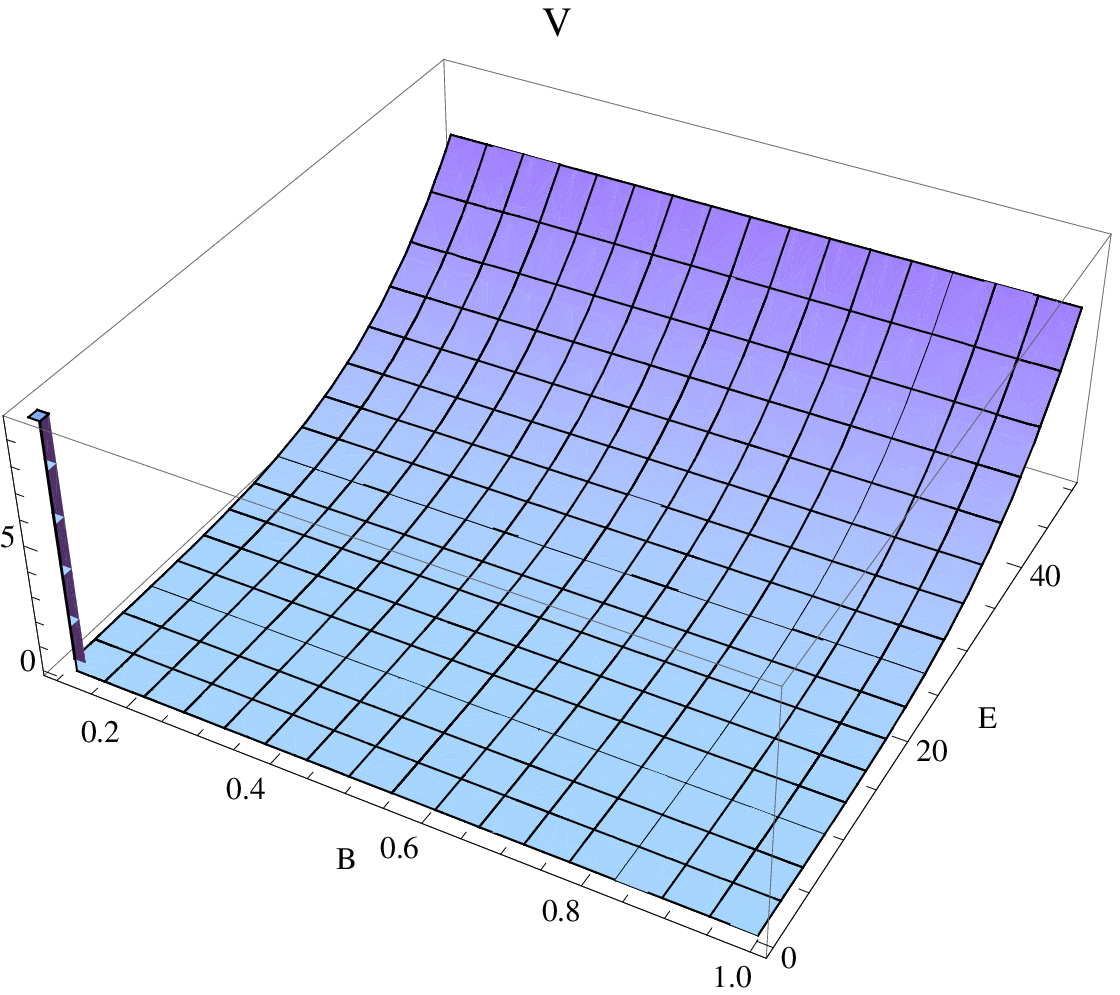}~~~~~~~~~~~~~~~~~~~~~~~~~~~\includegraphics[scale=0.6]{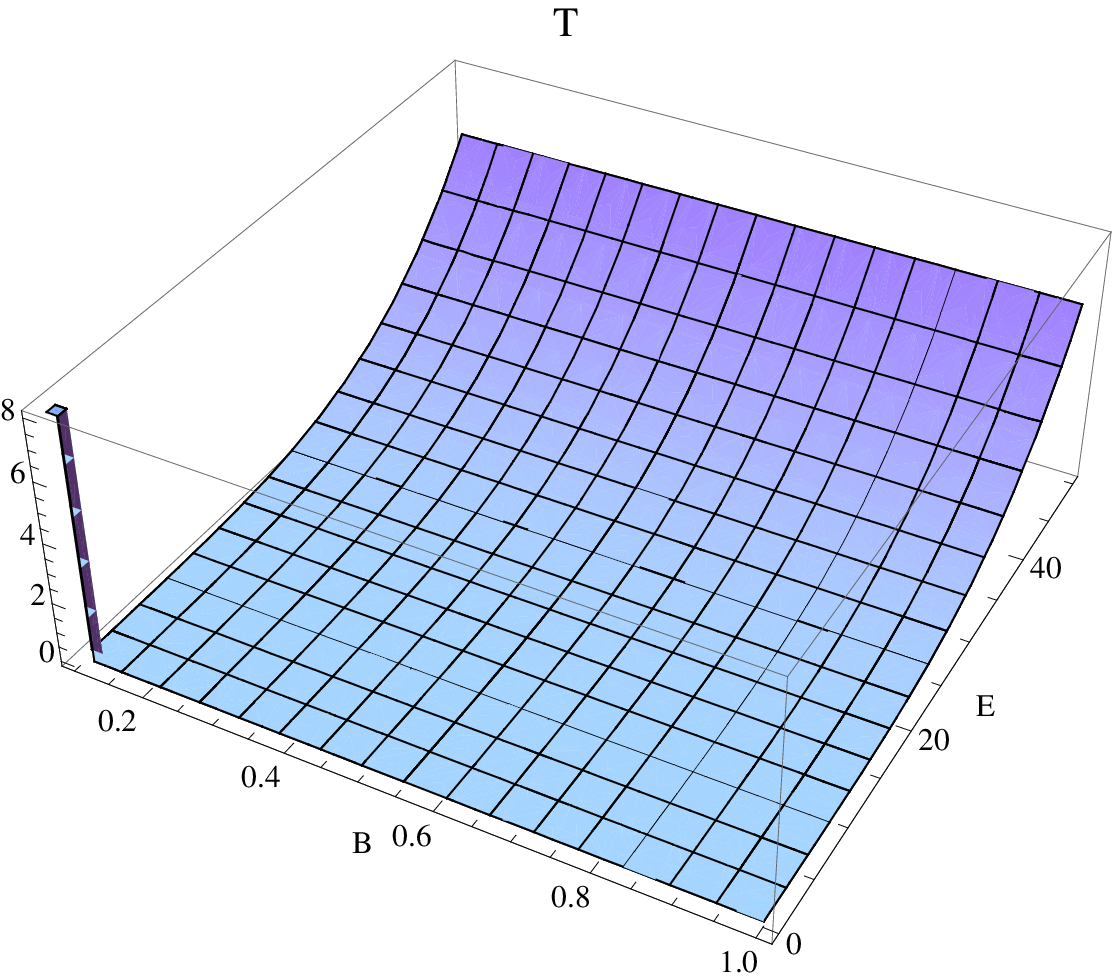}\\

\vspace{1mm} ~~~~~~Fig.2~~~~~~~~~~~~~~~~~~~~~~~~~~~~~~~~~~~~~~~~~~~~~~~~~~~~~~~~~~~~~~Fig.3~~~~~~~~\\

\vspace{6mm} Figs. 2 and 3 show the variation of $V$ and $T$
respectively against $B$ and $E$ for different values of
$~\alpha=1,\,\beta=2,\,m=6,\,n=5$ when a correspondence between EM
field and DBI-essence\\ scalar field is considered in Case I.
 \vspace{6mm}

\end{figure}

{\bf Case II:}   $\gamma \neq$ constant.\\

In this case let us assume $\gamma = \dot{\phi}^{s}$. So from
equation (23) we have $T(\phi) =
\frac{\dot{\phi}^{2s+2}}{\dot{\phi}^{2s}-1} > \dot{\phi}^{2}$.
Since $\gamma > 1$ so $\dot{\phi}^{s} > 1$. Let us also assume
$V(\phi) = T(\phi)$. In this case we have the expression for
$\phi$ and $V(\phi)(=T(\phi))$ for a particular case when $s=-4$
as :

\begin{equation}
\phi=\int\frac{3\sqrt{2}(B^{2}-E^{2})}{a\sqrt{\frac{-B^{4}+E^{4}+8(B^{2}-E^{2})^{2}(B^{2}+3E^{2})\alpha+\beta)(2(B^{2}-E^{2})^{2}(B^{2}+E^{2}
)(-1+16(B^{2}-E^{2})\alpha)+(-2B^{2}+E^{2})\beta)}{B^{2}-E^{2}}}}da
\end{equation}

\begin{equation}
V=T=\frac{(2(B^{2}-E^{2})^{2}(B^{2}+E^{2})(-1+16(B^{2}-E^{2})\alpha)(-2B^{2}+E^{2})\beta)^{3}}{27(B^{2}-E^{2})^{6}
(1-\frac{(2(B^{2}-E^{2})^{2}(B^{2}+E^{2})(-1+16(B^{2}-E^{2})\alpha)(-2B^{2}+E^{2})\beta)^{4}}{81(B^{2}-E^{2})^{8}})}
\end{equation}

Figures 2 and 3 show the variation of $V$ and $T$ respectively
against $B$ and $E$ for different values of parameters in the
solution of case I. From 3 dimensional figures, we have seen that
$V$ and $T$ always increase with $E$ increases but slightly
increases with increase in $B$. Similarly, figure 4 shows the
variation of $V$ or $T$ against $B$ and $E$ in the solution of
case II. In this case we see that $V$ or $T$ always
decreases with $E$ increases but slightly increases with increase in $B$.\\

\begin{figure}
\includegraphics[scale=0.6]{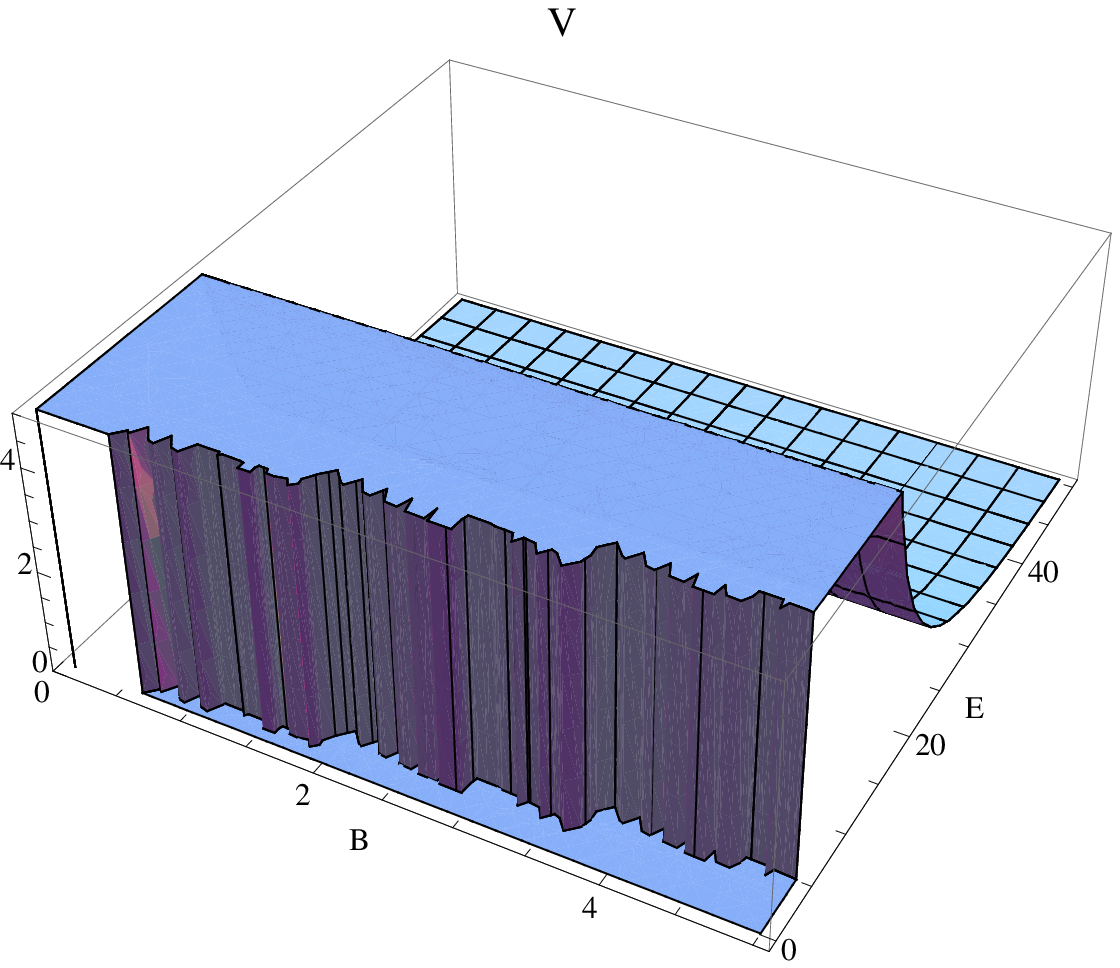}\\

\vspace{1mm} ~~~~~~~~~~~~Fig.4~~\\

\vspace{6mm} Fig. 4 shows the variation of $V$ or $T$ against $B$
and $E$ for different values of $~\alpha=2,\,\beta=3$ when a
correspondence between EM field and DBI-essence scalar field is
considered in Case II.
 \vspace{6mm}

\end{figure}

\begin{figure}
\includegraphics[scale=0.6]{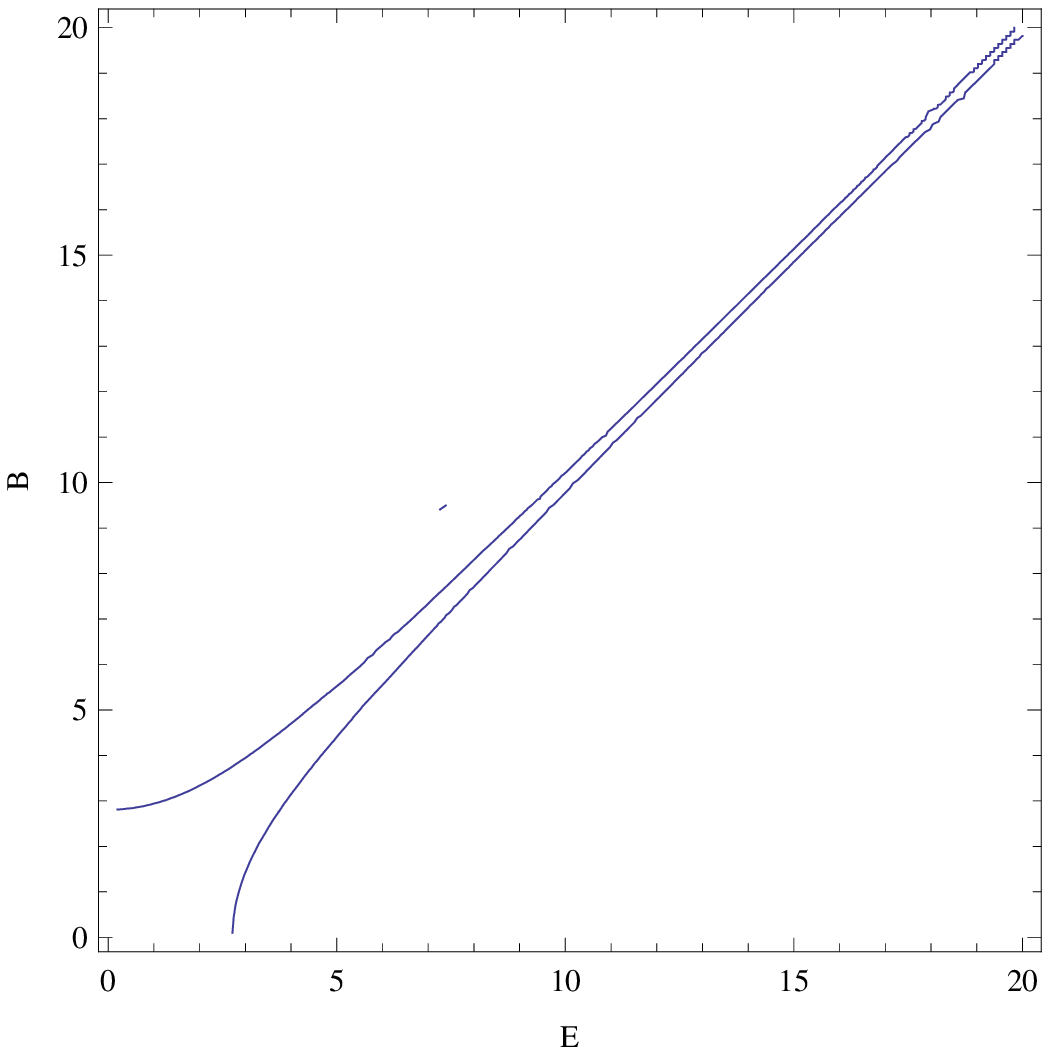}\\

\vspace{1mm} ~~~~~~~~~~~~~~~~~~~~~~Fig.5~~~~~~~~~~~\\

\vspace{6mm} Fig. 5 shows the variation of $B$ against $E$ for
different values of
$~\alpha=0.002,\,\beta=0.003,\,A=1/3,\,B_{1}=1,\,\alpha_{1}=1/3$
when a correspondence between EM field and Chaplygin gas is
considered.
 \vspace{6mm}

\end{figure}

\section{\bf{Correspondence between EM field and Chaplygin gas}}

A candidate for Q-matter is exotic type of fluid - the so called
Chaplygin gas which obeys the EOS [12]
$p_{ch}=-\frac{B_{1}}{\rho_{ch}}$ ($B_{1}>0$), where $p_{ch}$ and
$\rho_{ch}$ are respectively the pressure and energy density.
Later on the above equation was generalized to the form
$p_{ch}=-\frac{B_{1}}{\rho_{ch}^{\alpha_{1}}}~(0\leq\alpha_{1}\leq
1)$ and recently it was modified to the form
$p_{ch}=A\rho_{ch}-\frac{B_{1}}{\rho_{ch}^{\alpha_{1}}}$ ($A>0$),
which is known as Modified Chaplygin Gas [21].\\

Now comparing the energy density and pressure with electromagnetic
field we have a relation between $E$ and $B$ in implicit form as

\begin{eqnarray*}
\frac{1}{6}\left[8(-5+3A)B^{4}\alpha-8(1+9A)E^{4}\alpha+E^{2}\left(1-3A-\frac{(5+3A)\beta}{(B^{2}-E^{2})^{2}}\right)
+B^{2}\left(
1-3A+48E^{2}\alpha+48AE^{2}\alpha+\frac{(7+3A)\beta}{(B^{2}-E^{2})^{2}}\right)\right.
\end{eqnarray*}
\begin{equation}
\left.+3\times
2^{1+\alpha_{1}}B_{1}\left(B^{2}+E^{2}-8(B^{4}+2B^{2}E^{2}-3E^{4})\alpha+\frac{\beta}{-B^{2}+
E^{2}}\right)^{-\alpha_{1}}\right]=0
\end{equation}

The above expression is very complicated forms in $B$ and $E$. We
can not write $B$ in terms of $E$ or $E$ in terms of $B$
explicitly. So for some particular values of the parameters, we have plotted the
magnetic field $B$ against electric field $E$ in figure 5. From the figure, we see that
$B$ increases as $E$ increases. So in this scenarios, electro-magnetic field behaves
like modified Chaplygin gas model.\\

\section{\bf{Correspondence between EM field and Hessence dark energy}}

Wei et al [13] proposed a novel non-canonical complex scalar field
named ``hessence" which play the role of quintom. In the hessence
model the so-called internal motion $\dot{\theta}$ where $\theta$
is the internal degree of freedom of hessence plays a phantom like
role and the phantom divide transitions is also possible. The
pressure and energy density for the hessence model are given by

\begin{equation}
p_{h}=\frac{1}{2}(\dot{\phi}^{2}-\phi^{2}\dot{\theta}^{2})-V(\phi)
\end{equation}
and
\begin{equation}
\rho_{h}=\frac{1}{2}(\dot{\phi}^{2}-\phi^{2}\dot{\theta}^{2})+V(\phi)
\end{equation}
with
\begin{equation}
Q=a ^{3}\phi^{2}\dot{\theta}= \text{constant,}
\end{equation}

where $Q$ is the total conserved charge, $\phi$ is the hessence scalar field and $V$ is the
corresponding potential.\\

Comparing the energy density and pressure corresponding to EM
field and hessence field we have the expressions for scalar field
for the hessence and potential as

\begin{equation}
\phi=\int\frac{\sqrt{3}}{a}\sqrt{1+\frac{B^{2}+E^{2}+\frac{3(7B^{2}-5E^{2})}
{(B^{2}-E^{2})^{2}}-16(5B^{4}-6B^{2}E^{2}+E^{4})+\frac{6Q^{2}}{a^{6}\phi^{2}}}
{3(-16B^{4}+E^{2}+48E^{4}+B^{2}(1-32E^{2})-\frac{3}{B^{2}-E^{2}})}}da
\end{equation}
and
\begin{equation}
V=\frac{1}{6}\left(
B^{2}+E^{2}+8B^{4}\alpha-48B^{2}E^{2}\alpha+40E^{4}\alpha+\frac{-5B^{2}\beta+4E^{2}\beta}{(b^{2}-E^{2})^{2}}\right)
\end{equation}

In figure 6, we have drawn the hessence potential function $V$
against magnetic field $B$ and electric field $E$. From 3
dimensional figure, we have seen that $V$ always increases with
$E$ increases but slightly decreases with increase in $B$.\\

\begin{figure}
\includegraphics[scale=0.6]{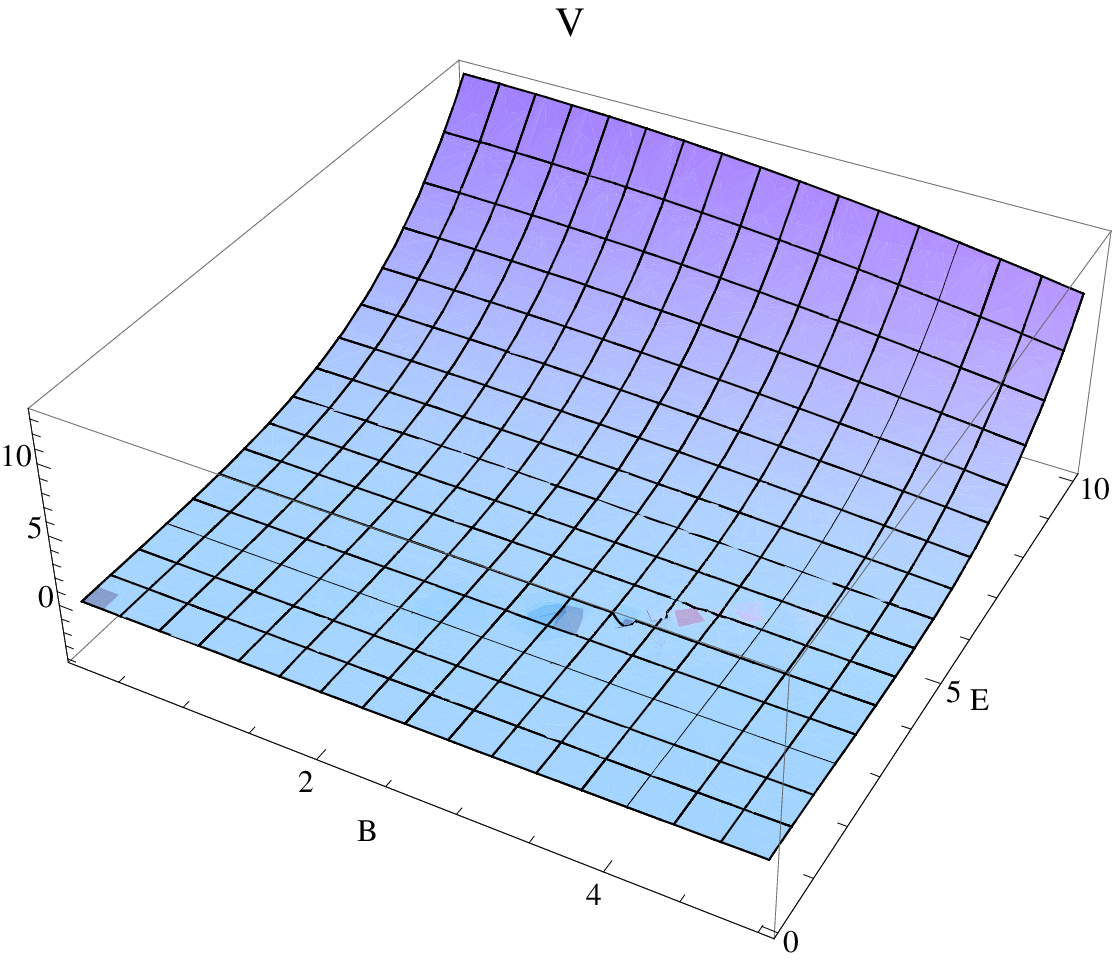}\\

\vspace{1mm} ~~~~~~~~~~~~~~~~~~~~~~Fig.6~~~~~~~~~~~\\

\vspace{6mm} Fig. 6 shows the variation of $V$  against $E$ and
$B$ for different values of $~\alpha=2,\,\beta=3$ when a
correspondence between EM field and hessence scalar field is
considered.
 \vspace{6mm}

\end{figure}

\section{\bf{Correspondence between EM field and k-essence}}

Another type of model responsible for late time acceleration
having non canonical kinetic term in Lagrangian called k-essence
which is originating form Born-Infeld string theory as a possible
model of inflation called k-inflation. The energy density and
pressure of k-essence scalar field $\phi$ are given by [14]

\begin{equation}
\rho_{k}=V(\phi)(\chi-3\chi^{2})
\end{equation}
and
\begin{equation}
p_{k}=V(\phi)(\chi-\chi^{2})
\end{equation}
where $\phi$ is the scalar field having kinetic energy $\chi=
\frac{1}{2}\dot{\phi}^{2}$ and $V(\phi)$ is
the k-essence potential.\\

Comparing the energy density and pressure corresponding to EM
field and k-essence field we have the expressions for k-essence
field and the potential as

\begin{equation}
\phi=\int\frac{2}{a}\sqrt{\frac{-(B^{2}-E^{2})^{3}(B^{2}+E^{2}+8(B{4}-6B^{2}E^{2}+5E^{4})\alpha)+(5B^{4}-9B^{2}E^{2}+4E^{4})\beta}
{(-B^{4}+E^{4}+8(B^{2}-E^{2})^{2}(B^{2}+3E^{2})\alpha+\beta)(16(B^{2}-E^{2})^{4}\alpha-4B^{2}\beta+3E^{2}\beta)}}da
\end{equation}
and
\begin{equation}
V=\frac{3(16(B^{2}-E^{2})^{4}\alpha-4B^{2}\beta+3E^{2}\beta)^{2}}{2(B^{2}-E^{2})^{4}(B^{2}+E^{2}+8(B{4}-6B^{2}E^{2}+5E^{4})\alpha)-2(5B^{2}
-4E^{2})(B^{2}-E^{2})^{2}\beta}
\end{equation}

In figure 7, we have drawn the k-essence potential function $V$
against magnetic field $B$ and electric field $E$. From 3
dimensional figure, we have seen that $V$ always increases with
$E$ increases but slightly increases with increase in $B$.\\

\begin{figure}
\includegraphics[scale=0.6]{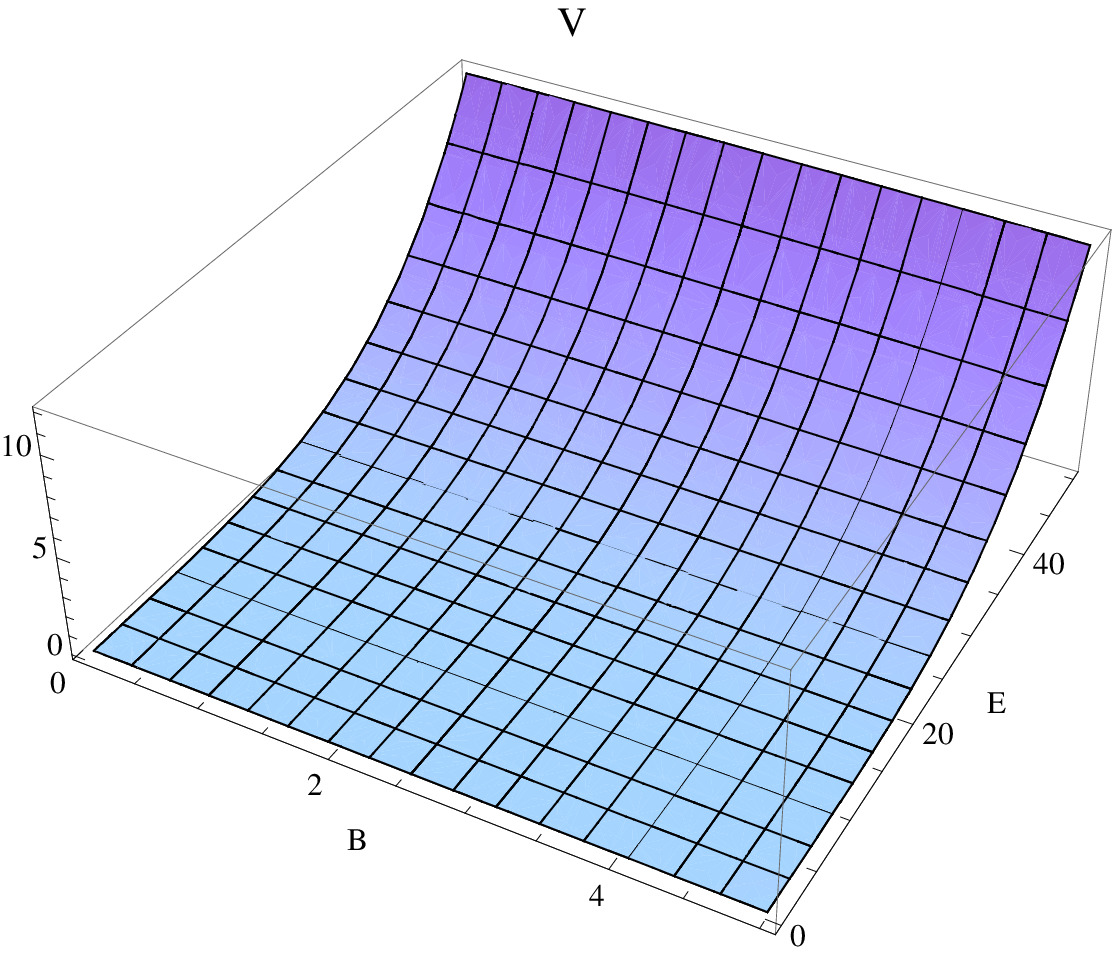}\\

\vspace{1mm} ~~~~~~~~~~~~~~~~~~~~~~Fig.7~~~~~~~~~~~\\

\vspace{6mm} Fig. 7 shows the variation of $V$  against $E$ and
$B$ for different values of $~\alpha=2,\,\beta=3$ when a
correspondence between EM field and k-essence scalar field is
considered.
 \vspace{6mm}

\end{figure}

\section{\bf{Correspondence between EM field and dilaton dark energy}}

String theory widely explain the present acceleration of the
universe and provides candidate for cold dark matter. We consider
here a dark energy model named dilaton dark energy which is
basically motivated from low energy limit of string theory.
Dilaton leads an initial inflationary phase followed by a kinetic
energy dominated phase. In the Einstein frame, the coefficient of
the kinematic term of the dilaton can be negative to behave as a
phantom-type scalar field.\\

The energy density and pressure of the dilaton dark energy model
are given by [15]

\begin{equation}
\rho_{d}=-X+Ce^{\lambda \phi}X^{2}
\end{equation}
and
\begin{equation}
p_{d}=-X+3Ce^{\lambda \phi}X^{2}
\end{equation}

where $\phi$ is the dilaton scalar field having kinetic energy
$X=\frac{1}{2}\dot\phi^{2}$, $\lambda$ is the characteristic
length which governs all non-gravitational interactions of the
dilaton and $C$ is a positive constant. Comparing the energy
density and pressure corresponding to EM field and dilaton dark
energy we have the expressions for scalar field as

\begin{equation}
\phi=\int\frac{1}{a}\sqrt{\frac{2(B^{2}-E^{2})^{2}(B^{2}+E^{2}+8(B^{4}-6B^{2}E^{2}+5E^{4})\alpha)-10B^{2}\beta+8E^{2}\beta}
{(B^{2}-E^{2})(-B^{4}+E^{4}+8(B^{2}-E^{2})^{2}(B^{2}+3E^{2})\alpha+\beta)}}da
\end{equation}

\section{\normalsize\bf{Discussions}}

In this work, we have considered the flat FRW universe filled with
electro-magnetic field. First, the Maxwell's electro-magnetic
field in linear form has been discussed and after that the
modified Lagrangian in non-linear form for accelerated universe
has been considered. The corresponding energy density and pressure
for non-linear electro-magnetic field have been calculated. We
have found the condition such that the electro-magnetic field
generates dark energy. Other types of dark energy candidates
namely tachyonic field, DBI-essence, Chaplygin gas, hessence dark
energy, k-essenece and dilaton dark energy have been discussed
shortly. The correspondence between the electro-magnetic field and
the above types of dark energies have been investigated. We have
reconstructed the potential functions and the scalar fields in
this scenario.\\

In the tachyonic field model, the potential function $V$ against
magnetic field $B$ and electric field $E$ have been drawn in
figure 1. From 3 dimensional figure, we have seen that $V$ always
increases with $E$ increases but slightly increases with the
increase of $B$. In DBI-essence model, figures 2 and 3 show the
variation of $V$ and $T$ respectively against $B$ and $E$ for
different values of parameters in the solution of case I. From
the figures, we have seen that $V$ and $T$ always increase with
$E$ increases but slightly increases with increase in $B$.
Similarly, figure 4 shows the variation of $V$ or $T$ against $B$
and $E$ in the solution of case II. In this case we see that $V$
or $T$ always decreases with $E$ increases but slightly increases
with increase
in $B$.\\

In modified Chaplygin gas model, we have found the relation
between electric field and magnetic field if the electro-magnetic
field behaves like modified Chaplygin gas equation of state. We
have found an implicit relation between $E$ and $B$. We have also
plotted the magnetic field $B$ against electric field $E$ in
figure 5. From the figure, we see that $B$ increases as $E$
increases. In hessence dark energy model, we have drawn the
hessence potential function $V$ against magnetic field $B$ and
electric field $E$ in figure 6 and have seen that $V$ always
increases with $E$ increases but slightly decreases with increase
in $B$. For k-essence dark energy model, the k-essence potential
function $V$ against magnetic field $B$ and electric field $E$
have been drawn in figure 7 and have seen that $V$ always
increases with $E$ increases but slightly increases with increase
in $B$. In every dark energy models, the expressions of scalar
fields have been calculated in terms of electric field and
magnetic field.\\\\

{\bf References:}\\

[1] A. G. Riess et al  \textit{Astron. J.} \textbf{116} 1009 (1998).\\\

[2] E. W. Kolb and M. S. Turner \textit{Addison-Wesley, Redwood City, CA} (1990).\\\

[3] W. de Sitter,  \textit{Proc. K. Ned. Akad. Wet.}, \textbf{19} 1217 (1917).\\\

[4] M. Novello and J. M. Salim,  \textit{Phys. Rev. D} \textbf{20} 377 (1979).\\\

[5] M. Novello et al, \textit{IJMPA} \textbf{1} 641 (1993).\\\

[6] G. L. Murphy,   \textit{Phys. Rev. D} \textbf{8} 4231 (1973).\\\

[7] R. Garc´ýa-Salcedo and N. Breton, \textit{IJMPA} \textbf{15} 4341 (2000).\\\

[8] H. Salazar, A. Garcia and J. Pleba$\acute{n}$ski  \textit{J. Math. Phys.} \textbf{28} 2171 (1987). \\\

[9] E. Ayo$\acute{n}$-Beato and A. Garcia, \textit{Phys. Rev. Lett.} \textbf{80} 5056 (1998).\\\

[10] A. Sen,  \textit{JHEP} \textbf{065} 0207 (2002).\\\

[11] J. Martin and M. Yamaguchi,  \textit{Phys. Rev. D} \textbf{77} 103508 (2008).\\\

[12] A. Kamenshchik et al, \textit{Phys. Lett. B} \textbf{511} 265 (2001).\\\

[13] H. Wei, R. G. Cai and D. F. Zhang,  \textit{class. Quant. Grav.} \textbf{22} 3189 (2005).\\\

[14] C. Armendariz-Picon et al, \textit{Phys. Rev. D} \textbf{63} 103510 (2001).\\\

[15] H. Q. Lu et al,  \textit{arXiv:} \textbf{0409309} [hep-th]  (2004).\\\

[16] C. S. Camara, M. R. de Garcia Maia, J. C. Carvalho and J. A.
S. Lima1, {\it Phys. Rev. D} {\bf 69} 123504 (2004).\\\

[17]  R. C. Tolman and P. Ehrenfest, \textit{Phys. Rev.}
\textbf{36}, 1791 (1930); M. Hindmarsh and A. Everett,
\textit{Phys. Rev. D} \textbf{58}, 103505 (1998).
\\\

[18] R.  Garc´ýa-Salcedo et al, \textit{arXiv:}
\textbf{1006.2274v1}
[gr-qc].\\\

[19]  M. Novello et al, \textit{IJMPA} \textbf{20} 2421 (2005).\\\

[20] U. Debnath and M. Jamil, arXiv:1102.1632 [physics.gen-ph].\\

[21] H. B. Benaoum, {\it hep-th}/0205140; U. Debnath, A. Banerjee
and S. Chakraborty, {\it Class. Quantum Grav.} {\bf 21} 5609 (2004).\\

\end{document}